\documentclass{svjour2} 
\smartqed  
\usepackage{graphicx}

\journalname{J Low Temp Phys (2010)}
\begin{document}

\title{The intrinsic features of the specific heat at half-filled Landau levels \\
of two-dimensional electron systems}

\author{Cristine Villagonzalo \and Rayda Gammag}

\institute{C. Villagonzalo \and R. Gammag \at
              Structure and Dynamics Group \\ 
              National Institute of Physics, University of the Philippines, \\
	Diliman, 1101 Quezon City, Philippines \\
              Tel.: +632-920-9749\\
              Fax: +632-928-0296\\
          \and
          C. Villagonzalo \at      
              \email{cvillagonzalo@nip.upd.edu.ph}           \\
           \and
          R. Gammag \at
              \email{rayda.gammag@up.edu.ph}
}

\date{Pre-print: "The final publication is available at www.springerlink.com." }

\maketitle

\begin{abstract}
The specific heat capacity of a two-dimensional electron gas is derived for
two types of the density of states, namely, the Dirac delta function spectrum 
and that based on a Gaussian function. For the first time, a closed form expression 
of the specific heat for each case is obtained at half-filling.
When the chemical potential is temperature-independent, 
the temperature is calculated at which the specific heat is a maximum. 
Here the effects of the broadening of the Landau levels are distinguished from
those of the different filling factors. In general, the results derived herein 
hold for any thermodynamic system having similar resonant states.
\end{abstract}

\section{Introduction}
\label{intro}

Electrons in a two-dimensional layer continue to be of interest in physics since 
they exhibit nonclassical behavior and are readily realizable in semiconductor 
heterostructures \cite{Davies98}. Nowadays research investigations on two-dimensional 
electron gas (2DEG) systems are aimed at understanding the plateaus in the quantum Hall conductance \cite{DasSarma97} and the de Haas - van Alphen effect or the magnetization oscillations \cite{Wilde06,Zhu03,Wang09}. The smooth drop of the plateaus and the 
extrema of the oscillations reveal broadened energy levels, that is, a finite density 
of states (DOS) \cite{Eisenstein85} between eigenvalues $E_n$ . These $E_n$, known as 
Landau levels, are the quantized energy spectrum ($E$) obtained when a strong magnetic 
field ($B$) is applied perpendicular to the system's plane \cite{Ando82} or in a tilted 
direction relative to it \cite{Ramos09}. The striking behavior of $E_n$ broadening, 
from which other exotic features arise, has been attributed to disorder due to 
the presence of impurities, defects and other inhomogeneities in the system
\cite{Xie90,Meinel01,Smith85}.

Ideally, the DOS of a noninteracting 2DEG in a perpendicular $B$ is given as a 
series of delta functions \cite{Ando82}, that is,
\begin{equation}
D(E) =  \frac{2eB}{h}\sum_n \delta(E-E_n)\;.
\label{eq:delta}
\end{equation}
Here  $e$ is the electron charge, $h$ is Planck's constant and 
$E_n = (n+1/2)\hbar \omega_c$ is the $n$th Landau energy level.
The latter depends on $B$ through the cyclotron frequency $\omega_c = eB/m^*$ for 
a given effective mass $m^*$. Such a DOS structure can be used to model actual 
materials with a very narrow distribution of energy carriers \cite{Wilde06}.

Usually, the actual shape of the density of states of a 2DEG is determined 
by making theoretical fits to the heat capacity data from experimental measurements. 
This procedure made on a 2DEG in GaAs-GaAlAs multilayers \cite{Gornik85}, 
for instance, yields a DOS with Gaussian peaks on a flat background. 
Many experimental and theoretical studies on the thermodynamic and magnetic 
properties of 2DEG systems, therefore, use a form of DOS based on a Gaussian function 
\cite{Ando82,Smith85,Zawadzki84,Zawadzki84b} such as
\begin{equation}
D(E) = \frac{2eB}{h} \sum_n \left(\frac{1}{2\pi}\right)^{1/2}\frac{1}{\Gamma} 
\exp\left[-\frac{(E-E_n)^2}{2\Gamma^2}\right]\;,
\label{eq:gaussian}
\end{equation}
where the broadening is taken into account by the parameter $\Gamma$.
 
In one study using this DOS, the authors of Ref.~\cite{Zawadzki84b} predicted 
that the specific heat and other magneto-thermal effects at high $B$ become 
universal functions of $k_B T/\Gamma$,  where $k_B$ is the Boltzmann's constant
and $T$ is the temperature. However, their numerical calculations at constant
electron concentration were unable to provide the exact dependence of these 
thermodynamic properties on $k_B T/\Gamma$.

Another study of an impurity-free quantum Hall system in a periodic rectangular 
geometry found that the specific heat at constant volume behaves as 
$C_V \propto \exp[-\Delta/T]/T^2$, where $\Delta$ is the energy gap between the 
ground state and the first excited state \cite{Chakraborty97}. The numerical calculations 
in Ref.~\cite{Chakraborty97} derived $C_V$ for different values of the filling 
factor $\nu$. This nominal number of filled Landau levels  is given as  $\nu=hN/Be$, 
where $N$ is the electron concentration. For various filling factors, their results 
yield a $C_V$ with a sharp peak at low $T$ whose height depends on $\Delta$.
However, their $C_V$ calculations show that finite size effects become dominant 
at low $T$. Without an exact expression for $C_V$, it would also be difficult to 
distinguish the $\exp[-\Delta/T]/T^2$ behavior from the dominant lattice contribution 
$\sim T^3$ at very low $T$. In general, since $C_V$ depends strongly on the nature 
of a given system such as the DOS, the exact behavior of $C_V$ as a consequence of its 
electronic structure is not easily determined.

In this work, the authors will show that a closed form expression of $C_V$ is obtained 
for a 2DEG system when the chemical potential $\mu$ does not vary with $T$. The latter 
condition is satisfied at a finite non-zero $T$ when the last occupied Landau level is 
half-filled, that is, the last occupied energy level is filled to half its degeneracy. 
Particular interest has been devoted to $\nu=1/2$ and other half-filled 
states since no plateau of the Hall resistance is observed unlike other fractional
$\nu$ with an odd denominator \cite{Stormer97}.  At half-filling, the chemical potential 
is right at the center of the Landau level where extended states are located \cite{DasSarma97}. 
There is an absence of a gap at this $\nu$ and some experiments have observed features 
of a Fermi-liquid state \cite{Jain09,Schulze07}.

Here a detailed derivation will be presented of $C_V$  at $\nu=n^{\prime}/2$ 
(where $n^{\prime}$ is an odd integer) for a spinless 2DEG having a delta-shaped or 
a Gaussian DOS in the presence of a perpendicular applied magnetic field. 
The exact relations for the specific heat as obtained for the first time in this work 
serve as a basis that characterizes the underlying density of states of an electron gas.
In addition, the condition at which the heat capacity is a maximum is established for 
both cases including the actual nature of its dependence on $\Gamma$. This will be 
particularly useful for the design of devices that are intended to resist changes in 
temperature. Lastly, remarks will also be given regarding $C_V$ at other filling factors
and hence, the general case will be formulated.

\section{The Temperature Dependence of the Chemical Potential}

\begin{figure}[t]
\vspace{1cm}
\centering
  \includegraphics[width=0.9\columnwidth,height=!]{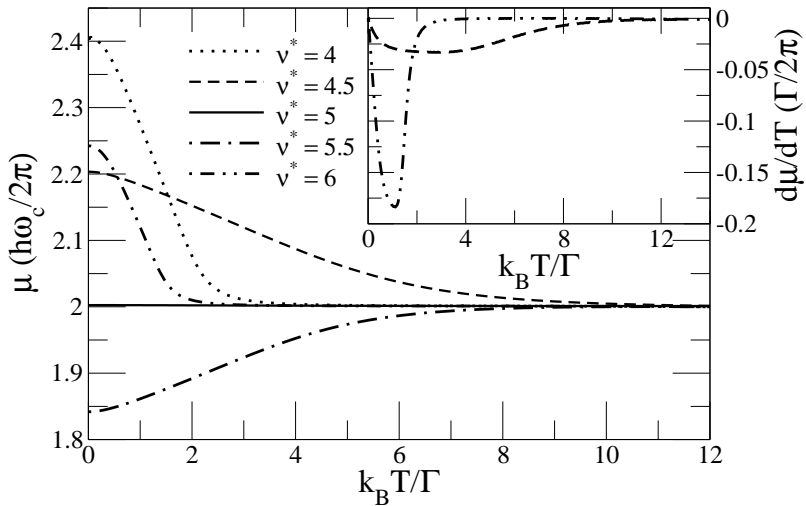}
\vspace{0.3cm}
\caption{The temperature-dependent chemical potential at various filling factors
for a 2DEG system with a Gaussian DOS. Here $\nu^{*}=5$ corresponds to half-filling 
for spinless electrons. This illustrates that $\mu$ does not change with $T$ in this case.
As examples, the inset provides the derivative of $\mu$ with respect to $T$ for $\nu^{*}=4.5$  
and $\nu^{*}=6$.}
\label{fig:mu}
\end{figure}

The chemical potential $\mu$, which measures the amount of energy 
required to add a particle in a given system, is generally a function of $T$ and $B$.
But at half-filling the $\mu$ is independent of $T$ since the 2DEG is metallic and
there is no energy gap between the conduction and the valence bands.
This is demonstrated in Fig.~\ref{fig:mu}. The chemical potential shown here 
is obtained for a fixed electron concentration, 
\begin{equation}
N = \int f(E) D(E)\;dE\;,
\end{equation}
using the root-finding method as outlined in Ref.~\cite{Gammag08}.
Here $f(E)$ is the Fermi-Dirac distribution function and it is given as
\begin{equation}
f(E)=\frac{1}{\exp[(E-\mu)/k_B T]+1}\;.
\end{equation}
The value of $N$ is set equal to  $3.6 \times 10^{11}$ cm$^{-2}$, which  
has the same order of magnitude as in experiments of 2DEG systems in semiconductors.
In addition, the 2DEG system used has a Gaussian DOS as given in 
Eq.~(\ref{eq:gaussian}) with $\Gamma=0.2$ meV. No spin-splitting is considered here.
Hence, in Fig.~\ref{fig:mu} the even integer $\nu^{*}$ corresponds to full filling 
and the odd integer $\nu^{*}$ to half-filling, that is, $\nu^{*}/2\rightarrow \nu$.
As the $T$ approaches zero, the $\mu$ of the fully-filled states ($\nu^{*}=4,\;6$)
varies within $\hbar \omega_c /2$ from the temperature-independent $\mu$ ($\nu^{*}=5$).
This is likewise observed for the non-half partially filled states
($\nu^{*}=4.5,\;5.5$), albeit at a gradual slope. This signifies that the $\mu$ behavior 
and, consequently, the specific heat nature of 2DEG systems at the non-half-filled states 
can be considered as deviations from their behavior at half-filling. 
In fact, the $\mu$ saturates to its half-filling value as $T$ increases. 
Treating the $\mu$ as $T$-independent is a valid approximation even for non-half-fillings
albeit limited only at high $T$.

\section{The Heat Capacity at Half-Filling}

The specific heat of a substance is a measure of the amount of energy needed 
to raise its temperature  by a degree. For an electron gas at constant volume, 
$C_V$ can be obtained from its internal energy $U$ as follows
\begin{equation}
C_V = \frac{\partial U}{\partial T} = \frac{\partial}{\partial T}
\int_{-\infty}^{\infty}f(E)(E-\mu)D(E)dE\;.
\label{eq:cv1}
\end{equation}

When the $\mu$ does not vary with $T$, only the $f(E)$ remains as the $T$-dependent 
function. The temperature derivative in Eq.~(\ref{eq:cv1}) then acts only on $f(E)$  
and Eq.~(\ref{eq:cv1}) can be rewritten as
\begin{equation}
C_V = k_B \int_{-\infty}^{\infty}
\frac{[(E-\mu)/2k_B T]^2}{\cosh^2[(E-\mu)/2k_B T]}D(E) \;dE\;.
\label{eq:cv2}
\end{equation}

\subsection{For a Delta-shaped DOS}

As shown in the numerical calculations of Ref.~\cite{Gammag08} at constant $N$,
despite the presence of the broadening of Landau levels, a characteristic
low temperature is reached at which the 2DEG approaches its ideal electron gas
behavior. Therefore, the continued use of a delta-shaped DOS  is justified 
in heat capacity measurements at strong $B$ and below a characteristic low $T$.

Substituting $D(E)$ of Eq.~(\ref{eq:delta}) in Eq.~(\ref{eq:cv2}), we obtain
the specific heat at half-filling for a delta-shaped DOS, that is,
\begin{equation}
C_V = k_B \frac{2eB}{h} \sum_n
\frac{[(E_n-\mu)/2k_B T]^2}{\cosh^2[(E_n-\mu)/2k_B T]} \;.
\label{eq:cv3}
\end{equation}
Note that the number of occupied Landau levels and the chemical potential are 
determined if the electron concentration is known. For this case, the contribution 
of the last occupied Landau level, $n=n_{max}$, to the $C_V$ is zero since $E_n=\mu$.

The specific heat expressed in Eq.~(\ref{eq:cv3}) exhibits a single peak at 
the low temperature region. The temperature at which $C_V$ is a maximum can be 
obtained by taking $\partial C_V/ \partial T = 0$ and performing a second-derivative 
test. The non-trivial condition at which this is satisfied is when
\begin{equation}
\sum_n \omega_n 
\frac{ \tanh(\omega_n) }{ \cosh^2(\omega_n) } \left[
\left( \omega_n \right) - \coth \left( \omega_n \right)
\right] = 0\;,
\label{eq:coth}
\end{equation}
where  $\omega_n = (E_n- \mu)/2k_B T$ and $E_n \neq \mu$. The maximum specific heat becomes
\begin{equation}
C_{V,\;{max}} = k_B \frac{2eB}{h} \sum_{n=0}^{n_{max}-1}
\frac{1}{\sinh^2[(E_n-\mu)/2k_B T]} \;.
\label{eq:cv3b}
\end{equation}

The value of $T$ when $C_V$  is a maximum, $T_{\mbox{\small peak}}$, can be determined at 
$n=n_{max}-1$ because this is the most dominant term. Also, 
$E_{n_{max}-1} \neq \mu$ and Eq.~(\ref{eq:coth}) remains valid. 
At this level, solving  for $\omega_n - \coth(\omega_n) = 0$ yields  $\omega_n = 1.19968$. 
Thus, we find that $T_{\mbox{\small peak}} \simeq 0.416778 (E_{n_{max}-1}-\mu)/k_B$. 
This result also provides a measure of the Landau level spacing.

The derivations herein are valid for any $B$ and for any $\mu(B)$ as long as the 
condition of half-filling is met or whenever the chemical potential does not change
with temperature. 

\subsection{For a Gaussian-shaped DOS}

There are other forms of the density of states used in the literature 
\cite{Wilde06,Zhu03,Gornik85} such as the Lorentzian, Gaussian with a constant 
background and the semielliptical distribution function. Here we consider only the 
Gaussian DOS as given in Eq.~(\ref{eq:gaussian}) to demonstrate the effect of 
a finite broadening $\Gamma$ as compared to the ideal impurity-free case. We keep 
in mind that in the limit as the $\Gamma \rightarrow 0$, the Gaussian DOS approaches 
the delta-shaped DOS as given in Eq.~(\ref{eq:delta}) since \\
$\delta(x) = \lim_{\Gamma\rightarrow 0} \;(1/2\pi\Gamma^2)^{1/2} \;\exp(-x^2/2\Gamma^2)$.

Substituting Eq.~(\ref{eq:gaussian}) in Eq.~(\ref{eq:cv2}) yields
\begin{eqnarray}
C_V = \frac{k_B}{\Gamma} \frac{2eB}{h} \left(\frac{1}{2\pi}\right)^{1/2} \sum_n  
\int_{-\infty}^{\infty} \exp\left( -\frac{(E-E_n)^2}{2\Gamma^2} \right) 
\nonumber \\
\times \frac{[(E-\mu)/2k_B T]^2}{\cosh^2[(E-\mu)/2k_B T]}
\;dE\;.
\label{eq:cv4}
\end{eqnarray}

For clarity and brevity, dimensionless energy parameters $\omega = (E-\mu)/2k_B T$ 
and $\omega_n=(E_n-\mu)/2k_BT$ are introduced such that the specific heat given 
in Eq.~(\ref{eq:cv4}) reduces to
\begin{eqnarray}
C_V = \frac{k_B}{\gamma} \frac{2eB}{h}\left(\frac{1}{\pi}\right)^{1/2}\sum_n
\int^{\infty}_{-\infty} \exp\left[-\frac{(\omega-\omega_n)^2}{\gamma^2}\right]
\times\left[\frac{\omega}{\cosh(\omega)}\right]^2  d\omega\;.
\label{eq:cv5}
\end{eqnarray}
Here $\gamma = (1/2)^{1/2} \Gamma/k_B T$ is the dimensionless broadening parameter.

The integral in the specific heat expression, Eq.~(\ref{eq:cv5}), is evaluated in 
two sequences of integration by parts which involve the integrals from 
Ref.~\cite{Gradshteyn07}: that is, 
(i) following Integral No.~8 of Section 2.477 \cite{Gradshteyn07} 
\begin{eqnarray}
\int \frac{\omega^2}{[\cosh(\omega)]^2} \;d\omega = \omega^2 \tanh(\omega) 
- 2\sum_{k=1}^{\infty} 
\frac{2^{2k}(2^{2k}-1)B_{2k}}{(2k+1)(2k)!}\omega^{2k+1}\;,
\end{eqnarray}
and that of (ii) Integral No.~7 of Section 2.479 \cite{Gradshteyn07}
\begin{equation}
\int \omega^p \tanh(\omega) \;d\omega = \sum_{k=1}^{\infty} 
\frac{2^{2k}(2^{2k}-1)B_{2k}}{(p+2k)(2k)!}\omega^{p+2k}\;,
\end{equation}
where $|\omega|<\pi/2$ and $p>-1$. The constant $B_{2k}$ is the $2k$-th Bernoulli 
number. The infinite integration of products of powers and hyperbolic functions 
are best treated using the Bernoulli series as a generating function in regions 
where the hyperbolic functions converge.

In terms of these integrals, the specific heat for a Gaussian DOS is given as
\newpage
\begin{eqnarray}
C_V = \frac{k_B}{\gamma} \frac{2eB}{h}\left(\frac{1}{\pi}\right)^{1/2}\sum_n
\left\{
\left[
\exp\left[-\frac{(\omega-\omega_n)^2}{\gamma^2}\right]\times
\right. \right. \;\;\;\;\;\;\;\;\;\;\;\;\;\;\;\;\;\;\;\;\;\;\;\;\;\;\;\;\;\;\;\;\;\;\;
\nonumber \\
\left(\omega^2 \tanh(\omega) 
+\sum_{k=1}^{\infty} \frac{2^{2k+1}(2^{2k}-1)B_{2k}}{(2k)!} \right. 
\;\;\;\;\;\;\;\;\;\;\;\;\;\;\;\;\;\;\;\;\;\;\;\;\;\;\;\;\;\;\;\;\;\;\;\;
\nonumber\\
\left. \left.
\times \left[ -\frac{\omega^{2k+1}}{2k+1}
+\frac{1}{\gamma^2}\frac{\omega^{2k+3}}{2k+3} 
-\frac{\omega_n}{\gamma^2}\frac{\omega^{2k+2}}{2k+2}
\right] 
\right)
\right]_{\omega=-\pi/2}^{\omega=\pi/2} 
\nonumber\\
+
\frac{1}{\gamma^2}
\sum_{k=1}^{\infty} \frac{2^{2k+2}(2^{2k}-1)B_{2k}}{(2k)!} 
\left[\int_{-\pi/2}^{\pi/2}
 \exp\left[-\frac{(\omega-\omega_n)^2}{\gamma^2}\right] 
\right.
(\omega-\omega_n)  \;\;\;\;\;\;\;\;\;\;\;\;\;\;
\nonumber \\
\left. \left.
\left( -\frac{\omega^{2k+1}}{2k+1}
-\omega_n\frac{\omega^{2k+2}}{2k+2}
+\frac{\omega^{2k+3}}{2k+3}
\right)
d\omega\right]\right\}\;. \;\;\;\;\;\;\;\;\;
\label{eq:cv6}
\end{eqnarray}
The remaining integrals in Eq.~(\ref{eq:cv6}) are of the form
\begin{eqnarray}
\int_{-\pi/2}^{\pi/2} \exp\left[-\frac{(\omega-\omega_n)^2}{\gamma^2} 
\right]\;\omega^s
\;d\omega= 
{\sum_{j=0}^{s}} 
\frac{s!}{j!(s-j)!}\omega_n^{s-j}\gamma^{j+1}\;F(j)\;,
\label{eq:gauss1}
\end{eqnarray}
where for the case when $j$ is odd
\begin{eqnarray}
F(j)=
-\frac{[(j-1)/2]!}{2}
 \exp\left(-\frac{(\omega-\omega_n)^2}{\gamma^2}\right)
\left.
\sum_{l=0}^{(j-1)/2} \frac{1}{l!}\left( \frac{\omega-\omega_n}{\gamma}
\right)^{2l} \right|_{\omega=-\pi/2}^{\omega=\pi/2} 
\end{eqnarray}
and for the case when $j$ is even
\begin{eqnarray}
F(j)=
\left\{-\exp\left[-\frac{(\omega-\omega_n)^2}{\gamma^2} \right] \right.
\sum_{l=0}^{j/2-1} \frac{(j-1)!!}{2^{j/2-l}(2l+1)!!} 
\times 
\nonumber \\
\left(\frac{\omega-\omega_n}{\gamma}\right)^{2l+1}
\left.
+ \frac{(j-1)!!}{2^{j/2+1}} (\pi)^{1/2}\mbox{erf}\left[
\frac{\omega-\omega_n}{\gamma}\right]
\right\}_{\omega=-\pi/2}^{\omega=\pi/2}
\;.
\end{eqnarray}
In this equation, $\mbox{erf}[z] = (2/\sqrt{\Pi})\int_0^z \exp[-z^2] \;dz$ 
is the error function. The series and integrals of Eq.~(\ref{eq:cv6}) converge. 
Therefore, one can determine the exact value of $C_V$ at a given $T$ and $B$ 
for a particular $\gamma$.

Note that, at half-filling, $E_n=\mu$ at the last occupied Landau level, $n=n_{max}$. 
In these cases, $\omega_n=0$. Hence, the integral of Eq.~(\ref{eq:gauss1})
vanishes. Therefore, the specific heat at the last occupied Landau level
given by Eq.~(\ref{eq:cv6}) evaluated from 
$\omega=-\pi/2$ to $\omega=\pi/2$ at $\omega_n=0$ yields 
\begin{eqnarray}
C_{V,\;n_{max}} = \frac{2eB}{h}\left(\frac{1}{\pi}\right)^{1/2}\frac{k_B}{\gamma}
\left\{
\exp\left[- \left(\frac{\pi}{2\gamma}\right)^2 \right] 
\right.\times
\nonumber \\
\left. \left(
\frac{1}{2}\pi^2\tanh\left(\frac{\pi}{2}\right)
+\frac{1}{\gamma^2}C_1-C_2 \right)\right\} \;,
\label{eq:cv7}
\end{eqnarray}
where the coefficients $C_1$ and $C_2$ are constants. From Eq.~(\ref{eq:cv6}),
these constant coefficients are derived to be as follows
\begin{equation}
C_1= \frac{1}{2}
\sum_{k=1}^{\infty} \frac{(2^{2k}-1)B_{2k}}{(2k+3)(2k)!}
\pi^{2k+3}\;,
\label{eq:c1}
\end{equation}
and
\begin{equation}
C_2= 2
\sum_{k=1}^{\infty} \frac{(2^{2k}-1)B_{2k}}{(2k+1)(2k)!}
\pi^{2k+1}\;.
\label{eq:c2}
\end{equation}
Their numerical values are $C_1 = 5.07023$ and $C_2 = 3.6412$.

Recall that $\gamma$ is inversely proportional to $T$. Setting to zero the 
derivative with respect to $T$ of Eq.~(\ref{eq:cv7}), a quartic equation is 
obtained. This yields one real root with $T>0$. 
This is the temperature at which the specific heat has a maximum at $E_n=\mu$, namely,
\begin{eqnarray}
T_{peak}= \frac{\Gamma}{k_B}\left\{-\frac{1}{8\pi^2C_1} \left(-12C_1-2\pi^2C_2+C_3 
\right. \right.  \;\;\;\;\;\;\;\;\;\;\;\;\;\;\;\;
\nonumber \\
-\left[144C_1^2+16\pi^2 C_1 C_2 +4\pi^4C_2^2 \right.
 \left. \left. \left.
-8C_1C_3-4\pi^2C_2C_3+ C_3^2
\right]^{1/2}\right)
\right\}^{1/2} ,
\label{eq:tpeak}
\end{eqnarray}
where $C_3=\pi^4\tanh[\pi/2] = 89.339$. This yields that 
$T_{\mbox{\small peak}} = 0.501815\;\Gamma/k_B$.

It is important to focus on the last occupied Landau level at $n=n_{max}$ at 
high $B$ and low $T$ because this is where the significant contribution to 
$C_V$ comes from \cite{Zawadzki84}. Since we are considering half-filling, 
$E_{n_{max}}=\mu$, the specific heat as derived from Eq.~(\ref{eq:cv4}) becomes
Eq.~(\ref{eq:cv7}). When this is rewritten in terms of $k_B T/\Gamma$, we find
that
\begin{eqnarray}
C_{V,\;n_{max}} = k_B \frac{2eB}{h}\left(\frac{1}{\pi}\right)^{1/2}
\exp\left[- \frac{\pi^2}{2}\left(\frac{k_B T}{\Gamma}\right)^2 \right]\times
\nonumber \\
\left(
2^{1/2}
\left[\frac{1}{2}\pi^2\tanh\left(\frac{\pi}{2}\right)-C_2\right]\frac{k_B T}{\Gamma}
+2^{3/2}C_1 \left[\frac{k_B T}{\Gamma}\right]^3 
\right)
\;,
\label{eq:cv8}
\end{eqnarray}
where the coefficients $C_1$ and $C_2$ are constants. They are given in 
Eq.~(\ref{eq:c1}) and Eq.~(\ref{eq:c2}), respectively. This result is consistent 
with the case for bulk metals that the $C_V \sim\;aT+bT^3$ where $a$ and $b$ are 
arbitrary coefficients. The linear in $T$ behavior of $C_V$ comes from the electronic
contribution, while the $T^3$ behavior is associated with lattice vibrations. 
This broadening of the Landau levels as represented by a Gaussian DOS that yields 
the $T^3$  contribution to the specific heat may be in part of the same nature 
as phonons. In addition to the similarity with 
metallic bulk properties, we obtain a Gaussian dependence on $k_B T/\Gamma$ for 
the specific heat as a natural consequence of the DOS, that is, 
$C_V\sim\exp[-(k_B T/\Gamma)^2]$. This is unlike in Ref.~\cite{Chakraborty97} 
where $C_V\sim \exp[-\Delta/T]$. Thus, in this work, the drop in $C_V$ after the 
$T_{\mbox{\small peak}}$ is caused by the Gaussian tail.

Equation (\ref{eq:cv8}) serves as a reference for the $C_V$ measurement as 
$\Gamma$ is varied. When $\Gamma$ is comparable to or greater than $k_B T$,  
$C_V$ decreases as $\Gamma$ increases for a given $T$.
This is due to the lesser energy requirement to move from one energy level to the 
next since there is a larger overlap between neighboring Landau levels.
When $\Gamma << k_B T$, the exponential behavior of Eq.~(\ref{eq:cv8}) dominates. 
In this case, $C_V$ increases as $\Gamma$ increases.
We have assumed here that $\Gamma$ is a constant for a given $B$ and $T$. 
Also, $\Gamma$ is assumed to be the same for all Landau levels. 
However, the effect of $\Gamma$ will be balanced by an increasing $B$ that increases the separation between Landau levels while raising the degeneracy of the states.
Even when a $B$-dependent critical value of $\Gamma$ is reached wherein the blend of states disappear \cite{Ramos09}, the expression for $C_{V,\;n_{max}}$  will still hold.

Furthermore, the functional form of $C_V$ in Eq.~(\ref{eq:cv8}) yields a 
single peak. As obtained from Eq.~(\ref{eq:tpeak}), this occurs at 
$T_{\mbox{\small peak}}=0.501815\;\Gamma/k_B$.
This linear proportionality with $T_{\mbox{\small peak}}$  and Eq.~(\ref{eq:cv8}) 
can be used to directly measure the broadening parameter of the DOS for a given 
$B$ whenever experimental data for the $C_V$ in a 2DEG system are available.

\section{A Note on the General Case}

When the chemical potential is temperature-dependent, the derivative of the Fermi
function is of the general form
\begin{equation}
\frac{\partial f}{\partial T} = -\frac{\partial f}{\partial E}
\left[\frac{E-\mu}{T} +\frac{\partial \mu}{\partial T}\right]\;.
\label{eq:dfdT2}
\end{equation}
This will yield two terms in the specific heat as given in 
Refs.~\cite{Zawadzki84,Zawadzki84b}.
But aside from $f(E)$, the specific heat also has an explicit dependence
on  $\partial \mu/\partial T$. Evaluating the temperature derivative
in Eq.~(\ref{eq:cv1}) results into
\begin{equation}
C_V =\int_{-\infty}^{\infty}\frac{\partial f(E)}{\partial T}(E-\mu)D(E)dE
-\int_{-\infty}^{\infty}f(E)\frac{\partial\mu}{\partial T}D(E)dE\;.
\label{eq:cv11}
\end{equation}
Using in this expression the Eq.~(\ref{eq:dfdT2}), we obtain the detailed
formula for $C_V$ when $\mu=\mu(T)$ as follows
\begin{eqnarray}
C_V = k_B \int_{-\infty}^{\infty}
\frac{[(E-\mu)/2k_B T]^2}{\cosh^2[(E-\mu)/2k_B T]}D(E) \;dE
\;\;\;\;\;\;\;\;\;\;\;\;\;\;\;\;\;\;\;\;
\nonumber \\
+\frac{\partial \mu}{\partial T} \left[ \frac{1}{2}
\int_{-\infty}^{\infty}
\frac{[(E-\mu)/2k_B T]}{\cosh^2[(E-\mu)/2k_B T]}D(E) \;dE
-\int_{-\infty}^{\infty} f(E)D(E)dE
\right]
\label{eq:cv12}
\end{eqnarray} 
This is the general equation for the specific heat and is valid regardless 
of the filling factor used.
Hence, we find that the correction to Eq.~(\ref{eq:cv2}) for $C_V$ will depend 
on how large $\partial \mu/\partial T$ is. In the case of a Gaussian-shaped DOS,
we have numerically shown that the contribution from $\partial \mu/\partial T$ 
for a fixed concentration is small. In the example given in the inset of 
Fig.~\ref{fig:mu}, this contribution is of the order of $0.01-0.1$ meV/K.

In the case of the delta-shaped density of states, substituting the DOS of 
Eq.~(\ref{eq:delta}), in Eq.~(\ref{eq:cv12}) yields
\begin{eqnarray}
C_V = k_B \frac{2eB}{h} \sum_n \left\{
\frac{\omega_n^2}{\cosh^2[\omega_n]} 
+\frac{1}{2k_B}\frac{\partial \mu}{\partial T} 
\left[ \frac{\omega_n}{\cosh^2[\omega_n]} 
 -\frac{2}{\exp[2\omega_n]+1}\right] \right\}
\label{eq:cv13}
\end{eqnarray}
where $\omega_n=(E_n-\mu)/2k_B T$. We get two corrections terms to Eq.~(\ref{eq:cv3}).
Their values depend again on the magnitude of $\partial \mu/\partial T$.

Though the focus of this work is  on the 2DEG, the  theoretical model as presented 
is also applicable  for any system under the grand canonical ensemble formulation with 
either a delta-shape or a Gaussian-broadened density of states.

\section{Conclusions}

The cardinal behavior of the specific heat of two-dimensional electron 
gas systems in an applied perpendicular magnetic field is derived 
for the cases wherein the density of states either has a delta-shape or 
a Gaussian form. This work establishes for both cases of the DOS  
the leading contributions to the specific heat and the temperature 
condition at which the $C_V$ is a maximum. 
The exact expressions obtained for $C_V$ hold when the chemical potential is 
independent of the temperature or the 2DEG is in a half-filling state. 
At low $T$ and when the $\mu$ depends on $T$, the formulation derived herein 
remains valid with an additional term in $C_V$ that is proportional to the 
$\partial \mu/\partial T$.
Furthermore, the results provide the linear proportionality of the broadening of 
the Landau levels to the temperature at which the $C_V$ attains its highest value. 
Experimental realizations of the results for the broadening parameter can be made,
for example, if it is induced by the presence of impurities, by varying impurity levels 
of a half-filled 2DEG system and monitoring the corresponding changes of the temperature 
at which the $C_V$ is a maximum.
When effects of various spin configurations are then considered, the derivations in
this work can be used to distinguish them from the effects of the temperature dependence
of the $\mu$ and the broadening of the Landau levels.

\begin{acknowledgements}
This work is supported by the Office of the Vice Chancellor for Research and 
Development (OVCRD) of the University of the Philippines Diliman under Project 
No.~080812 PNSE.
\end{acknowledgements}

\bibliographystyle{spphys}

\end{document}